# Phototaxis of synthetic microswimmers in optical landscapes

Celia Lozano[1,2], Borge ten Hagen[3], Hartmut Löwen[3] & Clemens Bechinger[1,2]

Many microorganisms, with phytoplankton and zooplankton as prominent examples, display phototactic behaviour, that is, the ability to perform directed motion within a light gradient. Here we experimentally demonstrate that sensing of light gradients can also be achieved in a system of synthetic photo-activated microparticles being exposed to an inhomogeneous laser field. We observe a strong orientational response of the particles because of diffusiophoretic torques, which in combination with an intensity-dependent particle motility eventually leads to phototaxis. Since the aligning torques saturate at high gradients, a strongly rectified particle motion is found even in periodic asymmetric intensity landscapes. Our results are in excellent agreement with numerical simulations of a minimal model and should similarly apply to other particle propulsion mechanisms. Because light fields can be easily adjusted in space and time, this also allows to extend our approach to dynamical environments.

[1] 2. Physikalisches Institut, Universität Stuttgart, D-70569 Stuttgart, Germany. [2] Max-Planck-Institut für Intelligente Systeme, D-70569 Stuttgart, Germany. [3] Institut für Theoretische Physik II: Weiche Materie, Heinrich-Heine-Universität Düsseldorf, D-40225 Düsseldorf, Germany. Correspondence and requests for materials should be addressed to C.B. (email: c.bechinger@physik.uni-stuttgart.de).

Despite their structural simplicity, active colloidal particles[1–5] exhibit many properties of motile microorganisms[6–9] including formation of clusters[10–14] and their response to gravitational[15,16] or flow[17] fields. Accordingly, active colloids provide an intriguing chance to understand the formation of dynamical structures in living systems[18–21] but may also find use as microrobots which—similar to their biological counterparts—autonomously navigate through complex environments[22]. In contrast to most previous experiments, where particles with spatially constant self-propulsion were considered[3,4,11,23–26], only little is known about synthetic systems with a position-dependent propulsion strength[27–30]. Such conditions, however, apply to many microorganisms, including bacteria and algae, which exhibit photo- or chemotactic motion allowing them to respond to external optical or chemical gradients[31–34]. In case of phototactic bacteria such as Rhodobacter sphaeroides[35], this is typically achieved by an intensity-dependent reorientation (tumbling) rate[33]; however, it should be mentioned that also other more sophisticated steering mechanisms are possible, for example, for the green alga Chlamydomonas[36] or the flagellate Euglena gracilis[37]. In general, a phototactic response in living systems requires complex internal feedback mechanisms between sensors and actuators[38], and it is therefore not obvious whether a similar behaviour can be realized with simple artificial microswimmers, which are not equipped with an elaborate internal network. Unlike living systems, the reorientation rate of colloidal particles is entirely determined by the rotational diffusion time $\tau_{rot} = 8\pi\eta R^3/k_B T$, which only depends on the particle radius $R$, the solvent viscosity $\eta$ and the thermal energy $k_B T$. However, if the reorientation rate does not respond to the local light intensity, no directed motion is possible. This is demonstrated in Fig. 1, where we show numerical simulations of the probability distribution $P(x)$ of self-propelled particles whose propulsion velocity varies linearly as a function of their position $x$. Although $P(x)$ asymmetrically broadens as a function of time, no macroscopic tactic particle flux is observed[28,39].

Here we experimentally demonstrate the orientational response and phototactic motion of spherical active colloids in a non-uniform light field. Owing to the coupling of both the particle's motility and its orientation to a light gradient, we observe particle transport towards lower intensities. Orientational particle response is achieved by breaking the axial symmetry of the velocity field around the particle, leading to an aligning torque. Because these torques saturate at higher gradients, we observe phototactic motion not only for monotonically increasing light fields, but also for periodic and asymmetric intensity profiles. Our experimental observations are in agreement with a theoretical model based on appropriate Langevin equations taking the position-dependent particle motility and orientational response into account. The latter is derived from the advective coupling of the heat flux in the illuminated particles to the solvent slip velocity at the particle surface. For a particle orientation misaligned with the intensity gradient, the theory predicts an aligning torque that saturates with increasing light intensity.

## Results

**Experimental characterization of the aligning torque.** Our experiments are performed with light-activated Janus particles, which are composed of optically transparent silica spheres (2.7 μm diameter) being capped on one side with a thin light-absorbing carbon layer. Upon illumination, such particles perform a self-diffusiophoretic active motion (with the cap pointing opposite to the direction of motion as being characteristic for polar particles) whose velocity $v_p$ is determined by the incident light intensity $I$ (refs 12,23; Fig. 2c). Under our experimental conditions, rotational and translational particle motion is limited to two dimensions because of hydrodynamic effects[40]. Periodic asymmetric intensity patterns are created by a laser line focus ($\lambda = 532$ nm) being scanned across the sample plane with 200 Hz. Synchronization of the scanning motion with the input voltage of an electro-optical modulator (EOM) leads to a quasi-static illumination landscape (Fig. 2b and Methods).

When we subject a diluted suspension of microswimmers to a one-dimensional (1D), linear intensity gradient $\nabla I$, we observe directed particle motion towards low intensities within a few seconds (Fig. 3a). We want to remark that optical gradient forces can be ruled out as the origin of the observed drift motion (Methods). After the light field is turned on at time $t_0$, particles rotate until their orientation $\hat{u}$ points antiparallel to the gradient. Since $\hat{u}$ is parallel to the self-propulsion velocity, this results in motion opposite to $\nabla I$. The reorientation is caused by breaking the axial symmetry of the slip velocity around the particle because of the inhomogeneous illumination, which leads to a viscous torque acting on the particle[41] (lower inset of Fig. 3b). To obtain further information about the phototactic response, we first investigated the reorientation dynamics for $t > t_0$. Therefore, we have studied the time evolution of $\theta(t)$, that is, the angle between $\nabla I$ and $\hat{u}$, for particles whose orientation at $t_0$ was $\theta = 0°$ (as shown in the upper inset of Fig. 3b). Within few seconds, particles orient antiparallel to $\nabla I$ (in our experiments, both clockwise and anticlockwise rotation is observed). The viscous torque $\mathbf{M}$ responsible for particle alignment is proportional to $\nabla I \times \hat{u}$, such that the noise-free angular velocity at low Reynolds

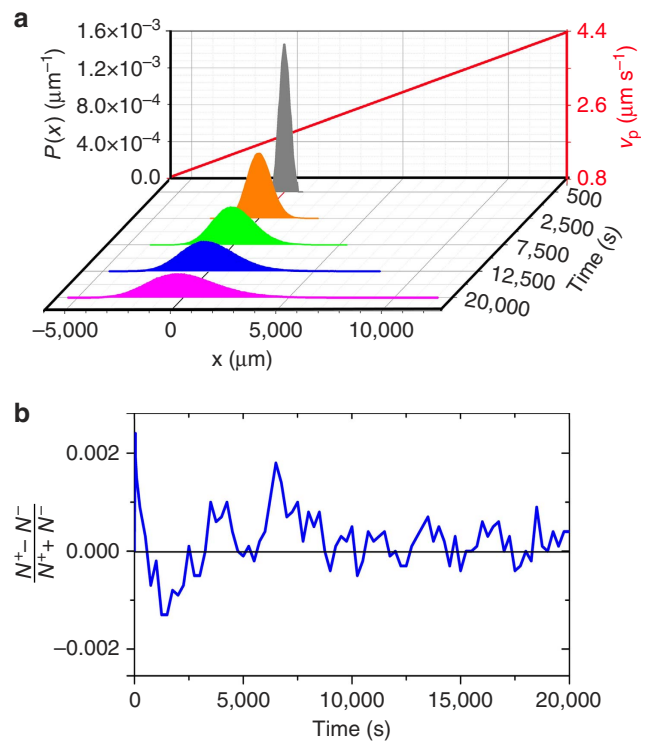

**Figure 1 | Particle motion with mere position-dependent propulsion velocity.** (**a**) Time evolution of the probability distribution $P(x)$ of self-propelled particles with variable self-propulsion velocity $v_p(x) = \tilde{v}(1 + x/a)$ for $x \geq -a$ as obtained from Brownian dynamics simulations (Methods). Although $P(x)$ becomes asymmetric at long times, no systematic drift of the particles to regions of higher motility is observed. (**b**) Time evolution of $(N^+ - N^-)/(N^+ + N^-)$, where $N^+$ and $N^-$ are the numbers of particles at $x > 0$ and $x < 0$, respectively. Obviously, a mere position-dependent motility does not induce a macroscopic particle current.

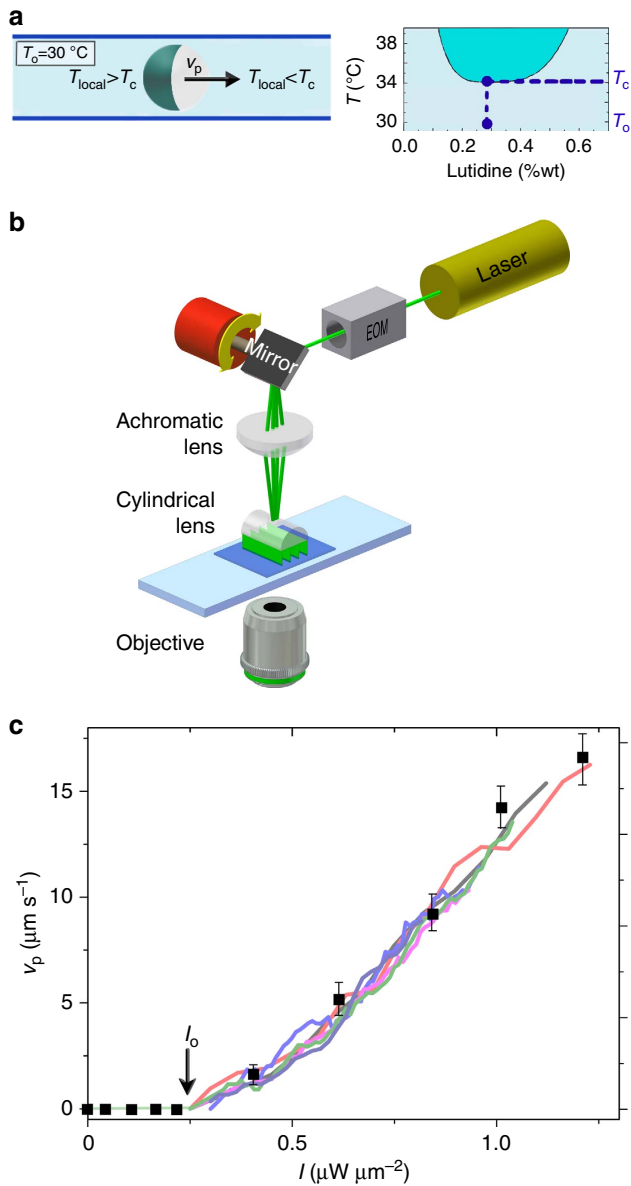

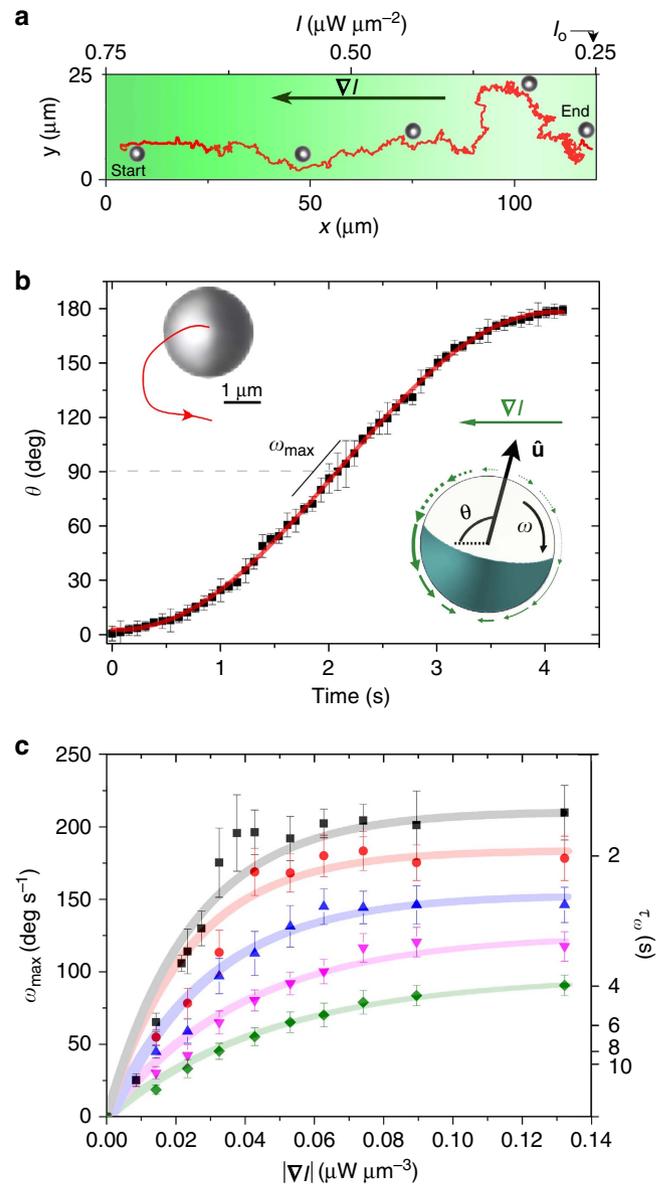

Figure 2 | Light-activated self-propulsion mechanism. (a) Sketch of the sample cell with a capped particle suspended in a binary critical mixture of water-2,6-lutidine. Illumination with light leads to heating of the cap and thus to local demixing resulting in active motion. (b) Experimental set-up for creation of periodic illumination landscapes by a scanned line focus of a laser beam (Methods). (c) Propulsion velocity $v_p$ versus illumination intensity $I$. Symbols (with error bars representing the s.d.) correspond to homogeneous illumination of the sample cell, while coloured lines were obtained in the presence of light profiles with different gradients ($|\nabla I| = 2 \times 10^{-3}\,\mu W\,\mu m^{-3}$ (pink), 0.027 $\mu W\,\mu m^{-3}$ (green), 0.037 $\mu W\,\mu m^{-3}$ (red), 0.115 $\mu W\,\mu m^{-3}$ (blue) and 0.156 $\mu W\,\mu m^{-3}$ (grey)). The agreement between the data for different intensity gradients demonstrates that $v_p$ is only determined by the local intensity incident on the particle.

number can be written as

$$\dot{\theta} = \omega_{max}\sin\theta, \quad (1)$$

where the amplitude $\omega_{max}$ depends on the intensity profile and sets the time $\tau_\omega$ needed to reorient the particle (for a mathematical definition of $\tau_\omega$ we refer to Methods). As shown in Fig. 3b (solid curve), the solution of the differential equation (1) (see Methods) is in excellent agreement with our experimental data for a prescribed and constant intensity gradient.

Figure 3 | Phototactic particle motion in constant light gradients. (a) Trajectory of an active particle in a gradient $|\nabla I| = 0.042\,\mu W\,\mu m^{-3}$. (b) Time evolution of the angle $\theta$ for $|\nabla I| = 0.02\,\mu W\,\mu m^{-3}$ and $I = 0.55\,\mu W\,\mu m^{-2}$. The data are averaged over 10 runs. Upper inset: snapshot of a particle and its trajectory (solid curve) during reorientation. Lower inset: sketch of an active colloid in a non-uniform light field with gradient $\nabla I$. The slip velocity (green arrows) becomes axially asymmetric, which results in an angular velocity $\omega$ (ref. 41). (c) Plot of the maximum angular velocity $\omega_{max}$ (left axis) and the corresponding reorientation time $\tau_\omega$ (right axis, see Methods for details) as a function of the gradient $|\nabla I|$ for different initial local intensities, that is, velocities ($I = 0.94\,\mu W\,\mu m^{-2}$ ($v_p = 12\,\mu m\,s^{-1}$), squares; $I = 0.69\,\mu W\,\mu m^{-2}$ ($v_p = 7\,\mu m\,s^{-1}$), circles; $I = 0.55\,\mu W\,\mu m^{-2}$ ($v_p = 5\,\mu m\,s^{-1}$), triangles; $I = 0.44\,\mu W\,\mu m^{-2}$ ($v_p = 3\,\mu m\,s^{-1}$), inverted triangles; $I = 0.35\,\mu W\,\mu m^{-2}$ ($v_p = 1.5\,\mu m\,s^{-1}$), diamonds). The error bars represent the s.d., and the solid curves show the theoretical fits (Methods).

We expect that the torque **M** depends on both the gradient $\nabla I$ and the absolute intensity at the particle position. To disentangle both effects, we varied the gradient of the intensity profiles and repeated the above measurements for particles at different initial positions $x_0 \equiv x(t_0)$ corresponding to different illumination intensities. The results for the maximum angular velocity $\omega_{max}$ and the corresponding reorientation time $\tau_\omega$ are shown in Fig. 3c, where each curve corresponds to $I(x_0) = $ const. Clearly, all curves saturate at large gradients. This nonlinearity can be explained theoretically by analysing the heat flux through the particle and its coupling to the solvent slip velocity at the particle surface (see Methods for details).

**Rectified motion.** Our results demonstrate that a phototactic motion requires a strong gradient and simultaneously a high intensity. For a monotonic spatial gradient, however, both conditions are met only within a narrow region. To achieve directed particle transport over arbitrarily long distances, we created periodic sawtooth-like light profiles (Fig. 4a). It should be emphasized that this is different compared with previous theoretical studies where a sawtooth-shaped potential[42,43], that is, a position-dependent drift force, was acting on a suspension of active particles[44–47]. In that case, the particle velocity is given by the superposition of the drift and the self-propulsion velocity vectors, which results in a particle current in the direction of the smaller potential gradient. In our case, however, no such external potential exists; it is only the propulsion velocity that is varied according to the asymmetric intensity profile. Hence, the underlying mechanism is different and the occurrence of a directed particle motion much less obvious.

Experimentally, sawtooth-shaped intensity profiles $I(x)$ have been created by application of a ratchet-shaped voltage to the electro-optical modulator (Fig. 2b). $I(x)$ is characterized by its period length $L$, the modulation amplitude $\Delta I$ and the length ratio of the two segments $a$ and $b$ (Fig. 4b). At the first glance, a particle current seems to be unlikely under such conditions because the opposite signs of $\nabla I$ in the two segments lead to opposite realigning torques that should—similar to the example trajectory shown in Fig. 3a—simply lead to an accumulation of the particles at the minima of $I(x)$. As a matter of fact, however, we observe a directed particle motion in $+x$ direction (see Supplementary Movie 1). This is demonstrated in Fig. 4c where we show typical trajectories $x(t)$ for different modulation amplitudes $\Delta I$ and $L = 33.5\,\mu m$, $a/b = 0.22$. Obviously, directed particle motion increases with increasing $\Delta I$. Because the orientation $\hat{\mathbf{u}}$ of a particle is parallel to its translational velocity, a net motion to the right implies that particles are able to pass segments $a$ with the 'wrong' orientation, that is, without being aligned opposite to $\nabla I$. This seems surprising because the realigning torques increase with $|\nabla I|$, which is larger in segments $a$ compared with $b$. To understand the occurrence of a rectified motion, one has to recall that the time required for the reorientation of a particle inside an intensity gradient, that is, $\tau_\omega$, decreases but rapidly saturates at large $|\nabla I|$ (Fig. 3c). The other relevant timescale is the particle's mean residence time $\tau_{r,i}$, that is, the time it spends in each segment $i$ ($i = a,b$) when travelling through the sawtooth-shaped light profile. When $\tau_{r,i} < \tau_{\omega,i}$, the particle will (on average) travel through the corresponding segment without a significant orientational response to the local gradient, which then leads to rectified motion.

In Fig. 4d, we exemplarily show the distribution of measured residence times in both segments for $\Delta I = 1.0\,\mu W\,\mu m^{-2}$ together with the corresponding reorientation times taken from Fig. 3c. Because $\tau_{\omega,i}$ depends not only on $|\nabla I|$ but also on the absolute intensity, we have plotted the values of $\tau_{\omega,i}$ for the mean intensity as vertical dashed lines in Fig. 4d. As can be seen, $\tau_{\omega,b}$ is smaller than the typical residence times $\tau_{r,b}$, which allows the particle to align opposite to the local gradient and thus leads to a motion to the right. This is in contrast to the situation in the segments $a$, where usually $\tau_{r,a} < \tau_{\omega,a}$. As a consequence, particles that travel from $b$ to $a$ segments are likely to maintain their previous direction of motion, which generates a particle current in the $+x$ direction. It should be emphasized that this rectification mechanism crucially depends on the saturation behaviour of $\tau_\omega$ versus $|\nabla I|$ (Fig. 3c). Without such saturation, the time $\tau_{\omega,a}$ would be much shorter, leading to $\tau_{r,a} > \tau_{\omega,a}$. As a consequence, particles would always align opposite to $\nabla I$, which results in a mere accumulation of the particles at the minima of the intensity profile. The corresponding time distributions for $\Delta I = 0.7$ $\mu W \mu m^{-2}$ are shown in Fig. 4e. In particular in the $a$ segments, we find an enhanced probability that $\tau_{r,a} > \tau_{\omega,a}$, which leads to events where the particle moves to the left (see inset Fig. 4c) and thus explains why the particle transport is strongly reduced under such conditions.

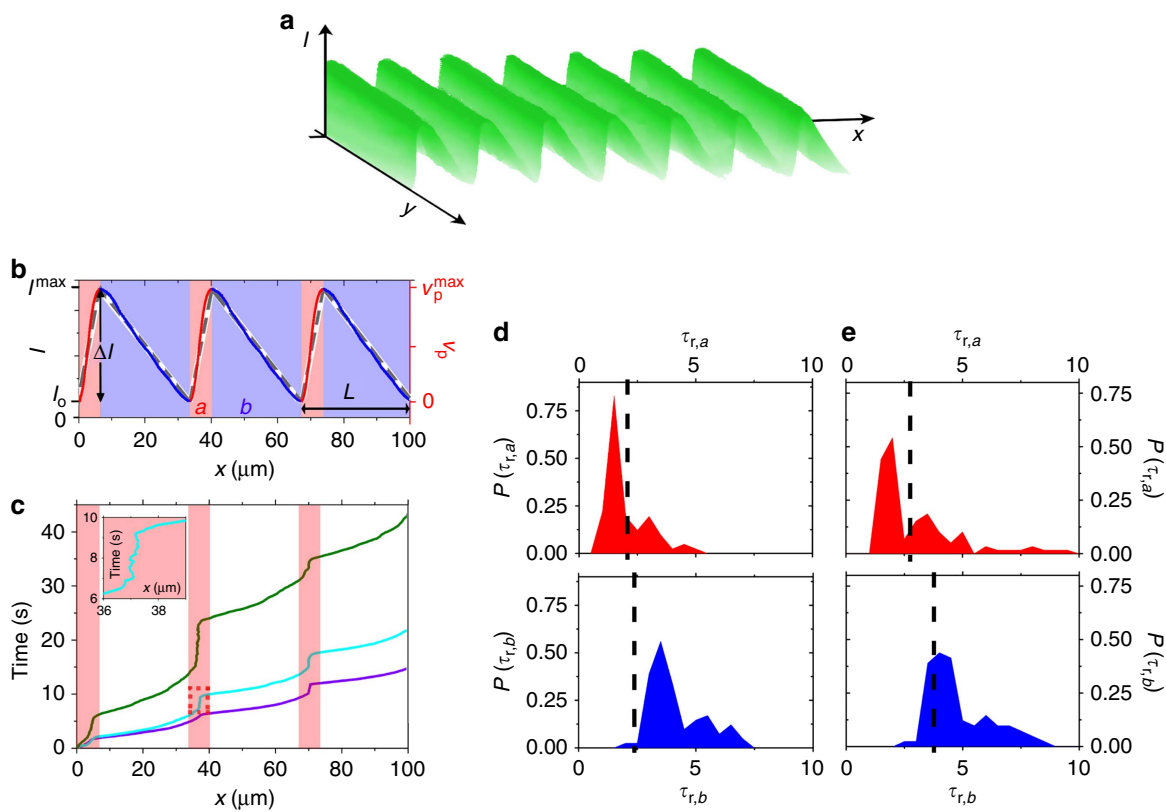

**Figure 4 | Rectification mechanism.** (**a**) Measured intensity profile of a one-dimensional asymmetric light field with period length $L = 33.5\,\mu m$. (**b**) Comparison of experimental (solid line) and numerical (dashed line) intensity profiles. The latter is approximated by segments with constant positive (*a*) and negative (*b*) gradients. (**c**) Particle trajectories for $a/b = 0.22$ and $\Delta I = 1.0\,\mu W\,\mu m^{-2}$ (purple), $0.70\,\mu W\,\mu m^{-2}$ (cyan) and $0.45\,\mu W\,\mu m^{-2}$ (green). Inset: magnification of the framed region demonstrating back and forth motion of a particle. (**d,e**) Probability distribution functions (PDFs) of $\tau_{r,i}$ ($i = a,b$) for $a/b = 0.22$ and (**d**) $\Delta I = 1.0\,\mu W\,\mu m^{-2}$ and (**e**) $\Delta I = 0.70\,\mu W\,\mu m^{-2}$, obtained from *a* segments (red) and *b* segments (blue), respectively. The vertical dashed lines indicate the corresponding mean values of $\tau_{\omega,i}$ taken from Fig. 3c.

The average velocity of the particles within one period can be calculated according to $\langle \dot{x} \rangle = \frac{1}{N} \sum_{j=1}^{N} \frac{L}{\tau_{r,j}}$, where $\tau_{r,j}$ is the time required to cross a single period $L$ and $N$ is the total number of periods. In Fig. 5a, we plotted $\langle \dot{x} \rangle$ versus $\Delta I$ for different ratios $a/b$. As expected, $\langle \dot{x} \rangle$ gradually decreases when $a/b \to 1$, that is, when the light landscape becomes more symmetric.

## Discussion

Our experimental findings demonstrate a strong rectified motion due to the coupling of the particle orientation to the gradient. In order to understand the above results in more detail, we performed numerical simulations where, in addition to the intensity-dependent particle velocity $\mathbf{v}_p = v_p(x)\hat{\mathbf{u}}$, the aligning torque $\mathbf{M}$ due to the light gradient was taken into account. The corresponding Langevin equations for the centre-of-mass position $\mathbf{r}(t)$ and the orientation $\hat{\mathbf{u}}(t)$ of the particle are

$$\dot{\mathbf{r}} = v_p(x)\hat{\mathbf{u}} + \zeta_{\mathbf{r}}, \quad (2)$$

$$\gamma_{rot}\dot{\hat{\mathbf{u}}} = (\mathbf{M} + \zeta_{\hat{\mathbf{u}}}) \times \hat{\mathbf{u}}, \quad (3)$$

where $\zeta_{\mathbf{r}}$ and $\zeta_{\hat{\mathbf{u}}}$ are translational and rotational noise terms (see Methods for details). The rotational damping constant $\gamma_{rot}$ sets the scale of the torque $M = |\mathbf{M}|$, such that $\gamma_{rot}/M$ determines the reorientation time $\tau_\omega$. To keep the theoretical model as simple as possible, we approximated the light profile and correspondingly $v_p(x)$ by two straight segments (dashed curve in Fig. 4b). The experimental findings show that the phoretic torque $\mathbf{M}$ aligning the particle antiparallel to the local intensity gradient depends both on the magnitude and the gradient of the illumination intensity (Fig. 3c). Employing scaling arguments, we use the expression

$$M = \gamma_{rot} v_p(x) g((\nabla I)^2) \sin\theta \quad (4)$$

with a saturating function $g((\nabla I)^2)$, which can be understood from the coupling of the heat flux through the particle to the solvent slip velocity. Further details are provided in Methods.

The numerical results (solid curves in Fig. 5a) are in good agreement with our data and correctly describe both the dependence of $\langle \dot{x} \rangle$ on the amplitude and the asymmetry of the light profile. In addition to a corroboration of our experimental results, the simulations highlight the importance of the nonlinear dependence of the restoring torque **M** (and the corresponding reorientation time $\tau_\omega$) on the light gradient (cf. Fig. 3c). In case of a linear relationship $\mathbf{M} \propto \boldsymbol{\nabla} I$, our simulations do not lead to a significant rectified motion, and particles mainly accumulate at the minima of $I(x)$ (see Methods for details).

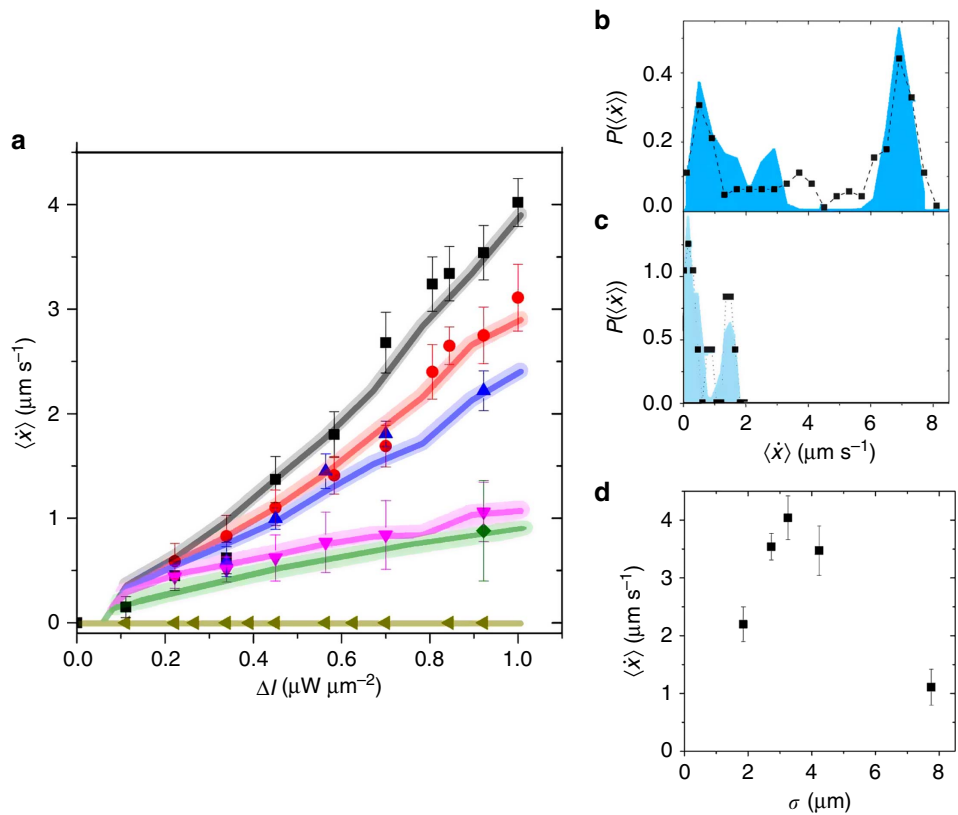

**Figure 5 | Particle current in periodic asymmetric light profiles.** (**a**) Average velocity $\langle \dot{x} \rangle$ versus intensity amplitude $\Delta I$ obtained from experiments (symbols) and numerical simulations (solid curves), for $a/b = 0.22$ (squares), 0.28 (circles), 0.33 (triangles), 0.55 (diamonds), 0.75 (inverted triangles) and 1.0 (pentagons). The experimental data points were obtained by averaging the velocity over ~200 periods each. Error bars correspond to 95% of the confidence interval. (**b,c**) PDFs of $\langle \dot{x} \rangle$ for $a/b = 0.22$ and (**b**) $\Delta I = 1.0 \mu W \mu m^{-2}$ and (**c**) $\Delta I = 0.22 \mu W \mu m^{-2}$, obtained from experiments (symbols) and numerical simulations (filled areas). (**d**) Average velocity $\langle \dot{x} \rangle$ for different particle diameters $\sigma$, for $a/b = 0.22$ and $\Delta I = 0.92 \mu W \mu m^{-2}$.

The particle alignment and the mean particle velocity are also influenced by thermal noise. This is shown by the velocity probability distribution functions, which are plotted in Fig. 5b,c for large and small light amplitudes $\Delta I$, respectively. The right peak of the bimodal distribution in Fig. 5b corresponds to particles that travel rather straight through the light landscape. This is only possible for particles whose orientation perfectly points in the $+x$ direction ($\theta \cong 0°$ in the $a$ segments). Under such conditions no torques arise and the particles travel continuously to the right. When reducing $\Delta I$, the influence of thermal fluctuations increases, which leads to deviations from a perfect particle alignment and thus to torques. This results in a back and forth motion around the minima of $I(x)$ due to opposite torques at those regions (inset Fig. 4c). Consequently, the probability of small $\langle \dot{x} \rangle$ is enhanced (Fig. 5c). We have also investigated the dependence of $\langle \dot{x} \rangle$ on the particle diameter $\sigma$. With decreasing $\sigma$, the phoretic torques become smaller[41], while for larger $\sigma$ the spatial averaging of the particle over the light landscape effectively reduces the amplitude $\Delta I$. This leads to a pronounced size dependence of $\langle \dot{x} \rangle$ (Fig. 5d).

Our findings demonstrate that a phototactic response can be observed in artificial microswimmers, which fortifies the resemblance with biological systems. Compared with microorganisms, where the underlying active reorientation is typically achieved by complex internal feedback loops, a tactic response to external light gradients can be accomplished in active colloids by breaking the axial symmetry of the diffusiophoretic slip velocity around the particle, which leads to particle alignment and directed motion along a light gradient. Because similar torques as discussed here are also expected for catalytically[26,48–50] or thermophoretically[4,51–53] self-propelled polar particles in concentration or temperature gradients[27,41,54], a comparable tactic behaviour is likely to occur also in other systems. Compared with topographical structures or external ratchet potentials, where particle currents havebeen experimentally demonstrated[44,45,47,55,56], the use of optical landscapes allows for dynamical adjustments because they can be easily varied in space and time. This offers promising opportunities for the navigation of active particles in dynamical environments.

## Methods

**Experimental details.** Artificial microswimmers were made from half-coating spherical silica particles (diameter $\sigma = 2.7\,\mu m$) with 20 nm carbon caps. Such particles are suspended in a binary critical mixture of water–2,6-lutidine with a lower critical point at $T_c = 34.1\,°C$ (phase diagram shown in Fig. 2a). The entire sample cell is kept constant at a bath temperature $T_0 = 30\,°C$. When illuminating the sample cell with light, which is only absorbed by the capped region, the solvent locally demixes. This leads to an inhomogeneous concentration field around the particle, which results in light-controlled active particle motion due to self-diffusiophoresis[23,57,58]. Only when the intensity exceeds a threshold value $I_0$ (see arrow in Fig. 2c), the cap is heated above $T_c$ and the particle exhibits self-propulsion. Below $I_0$, mere Brownian motion is observed.

Creation of periodic, asymmetric intensity patterns is achieved by a line focus of a laser beam ($\lambda = 532$ nm; half widths of 1 and 2,000 μm, respectively), which is rapidly scanned forth and back within the sample plane with a frequency of 200 Hz. At such frequencies the light pattern can be considered to be quasi-static. Images of the particles were acquired using video microscopy with a frame rate of 13 fps. From these images, the positions and orientations of particles were obtained using the Matlab image analysis software.

To ensure that directed particle motion in our experiments is not influenced by additional optical gradient forces, we have repeated all measurements at a bath temperature of 20 °C. Under such conditions the light field is not strong enough to induce self-propulsion, and only Brownian motion accompanied by a homogeneous particle distribution is observed.

**Orientational dynamics.** The active reorientation dynamics of a Janus particle in an intensity gradient $\boldsymbol{\nabla} I$ is governed by the differential equation (1) for the angle $\theta$ between $\boldsymbol{\nabla} I$ and the particle orientation $\hat{\mathbf{u}}$ (with $0° < \theta < 180°$) if thermal noise is neglected. Solving this equation gives

$$\cos\theta(t) = \tanh(\omega_{\max}(\bar{t} - t)), \quad (5)$$

where $\bar{t}$ is the time when the particle orientation is perpendicular to $\boldsymbol{\nabla} I$, that is, $\theta(\bar{t}) = 90°$. The result in equation (5) with $\omega_{\max}$ as the only fitting parameter was used to obtain the theoretical fit in Fig. 3b. In a corresponding way, all experimental data points in Fig. 3c were extracted from the measurements.

For a given value of $\omega_{\max}$, the reorientation time $\tau_\omega$ for a rotation from an angle $\varepsilon$ to an angle $\pi - \varepsilon$ is obtained as

$$\tau_\omega = -\frac{2}{\omega_{\max}} \ln\left(\frac{\cos(\pi - \varepsilon) + 1}{\sin(\pi - \varepsilon)}\right). \quad (6)$$

According to this result, $\tau_\omega$ becomes infinity for $\varepsilon = 0$. In that special case, the torque acting on the particle is zero. Thus, an alignment in the direction of the negative intensity gradient can only be achieved for an initial orientational fluctuation because of rotational Brownian motion. The values of $\tau_\omega$ on the right axis of Fig. 3c and the vertical dashed lines in Fig. 4d,e are obtained for $\varepsilon = 5°$.

**Brownian dynamics simulations in optical landscapes.** Most of the numerical results presented in this paper were obtained by Brownian dynamics simulations based on the overdamped equations of motion (2) and (3) for the time-dependent centre-of-mass position $\mathbf{r} = (x(t), y(t))$ and the orientation $\hat{\mathbf{u}} = (\cos\varphi, \sin\varphi)$ of a particle. Here $\varphi$ is the angle between the positive $x$ axis and the particle orientation, measured in the anticlockwise direction. Because the particle motion is effectively two-dimensional in the experiments as a consequence of hydrodynamic effects, that is, Brownian rotational quenching[40], displacements in vertical direction can be neglected. Thus, the vectorial rotational Langevin equation (3) can be written as a single equation for the angle $\varphi$ (ref. 59) as

$$\dot{\varphi} = \omega(x, \varphi) + \zeta_\varphi \quad (7)$$

with the angular velocity $\omega(x,\varphi)$. Brownian fluctuations are included in the equations of motion (equations (2) and (7)) by means of zero-mean Gaussian noise terms $\boldsymbol{\zeta}_\mathbf{r}$ and $\zeta_\varphi$ defined by the variances $\langle \boldsymbol{\zeta}_\mathbf{r}(t_1) \otimes \boldsymbol{\zeta}_\mathbf{r}(t_2) \rangle = 2 D_{\text{tr}} \mathbb{1} \delta(t_1 - t_2)$ and $\langle \zeta_\varphi(t_1) \zeta_\varphi(t_2) \rangle = 2 D_{\text{rot}} \delta(t_1 - t_2)$, where $\otimes$ denotes the dyadic product, $\mathbb{1}$ is the unit tensor, and $D_{\text{tr}}$ and $D_{\text{rot}}$ are the translational and rotational diffusion coefficients of a spherical Janus particle, respectively.

The position-dependent velocity profile $v_p(x)$ used in the simulations consists of two line segments with different slopes (dashed curve in Fig. 4b). The maximum value $v_{\max}$ of this sawtooth-like profile corresponds to the velocities obtained from the experiments (Fig. 2c). Moreover, we considered an offset $v_{\min} = 0.1 v_{\max}$, which takes the effects of the finite particle size into account. While in the theoretical model $v_p(x)$ is defined for every single point in space, a real particle always averages over the intensity profile in the vicinity of its surface. Thus, the active translational velocity is non-zero everywhere, which is incorporated in the simulations by the offset.

The angular velocity $\omega(x,\varphi)$ is directly related to the phoretic torque $M$ in equation (4) via $\omega = M/\gamma_{\text{rot}}$; thus,

$$\omega(x, \varphi) = \omega_{\max}(x) \sin\varphi = v_p(x) f\left(v'_p(x)\right) \sin\varphi \quad (8)$$

with a function $f$ being identical to the previously defined function $g((\boldsymbol{\nabla} I)^2)$, except for the different definition of the angle ($\varphi$ instead of $\theta$) and the direct dependence on the velocity gradient $v'_p(x) \equiv dv_p(x)/dx$ instead of the intensity gradient. The proportionality between the angular velocity $\omega$ and the linear velocity $v_p$ is illustrated by the following scaling argument: both quantities are obtained from the explicit slip velocity profile by appropriate integrations over the surface of the particle[27,60]. If at fixed intensity gradient all slip velocities are scaled by a factor $k$, this will just lead to a prefactor $k$ of the integrals for the total translational and angular velocities. Thus, the scaling in both cases is the same. In our model, this is ensured by the linear relation between $\omega$ and $v_p$ (see equation (8)). Finally, the function $f(v'_p(x))$ has to take the saturating behaviour observed in the experiments for the phoretic torque at high gradients into account. For the numerical simulations $f$ is approximated by

$$f\left(v'_p(x)\right) = \text{sgn}\left(v'_p(x)\right) \frac{C_1}{R} \left(1 - \exp\left(-C_2 \left|v'_p(x)\right|\right)\right), \quad (9)$$

where $R$ is the particle radius, and $C_1$ and $C_2$ are constants obtained by fitting the experimental data in Fig. 3c. A detailed theoretical explanation of the saturating behaviour is provided in the subsequent section.

**Explanation of the saturation.** To understand the saturating behaviour of the aligning torque with an increasing light intensity from a theoretical point of view, we analyse the heat flux in an illuminated Janus particle. For that purpose, we use an effectively 1D model for the temperature profile $T(x,t)$ in the coated region of the particle. The corresponding heat equation is given by

$$\frac{\partial T}{\partial t} = D_T \frac{\partial^2 T}{\partial x^2} + \alpha(I_0 + xI') - \gamma(T - T_0), \quad (10)$$

where the second term on the right hand side is a source term due to absorption of the incident light with absorption coefficient $\alpha$ and the last term with coefficient $\gamma$ represents the heat transfer from the cap to the bulk of the particle and to the solvent along the lateral part of the cap. The heat diffusion through the cap is determined by the thermal diffusivity $D_T$, and $T_0$ is the bulk temperature of the solvent. Finally, $I_0 + xI'$ is the considered linear light intensity profile with a constant gradient $I' \equiv |\nabla I|$ and a reference intensity $I_0$ at $x = 0$.

To determine the temperature profile inside the particle cap, we consider the stationary state, that is, $\partial T/\partial t = 0$. In that case, the solution of equation (10) is given by

$$T(x) = B_1 \exp(\omega x) + B_2 \exp(-\omega x) + \frac{\alpha}{\gamma}(I_0 + xI') + T_0 \quad (11)$$

with $\omega = \sqrt{\gamma/D_T}$ and constants $B_1$ and $B_2$, which have to be determined by choosing appropriate boundary conditions. These are defined by the coupling of the heat flux through the boundary of the particle cap with the solvent velocity at the particle surface (Fig. 6a,b). The heat flux inside the particle cap is determined by the local temperature gradient $T'(x)$ and is given by

$$j_Q = -C_c \rho_c D_T T'(x) = -\kappa T'(x) \quad (12)$$

with the specific heat capacity $C_c$, the mass density $\rho_c$ and the thermal conductivity $\kappa$ of the cap. On the other hand, the heat flux through the faces of the cap in the effectively 1D model (Fig. 6b) can be written as

$$j_Q = \frac{\Delta Q}{A_\perp \Delta t}, \quad (13)$$

where $A_\perp$ is the corresponding surface area. We consider a rectangular area with side lengths $d$ and $h$, thus $A_\perp = dh$. The heat $\Delta Q$ required to raise the temperature of a solvent volume $\Delta V$ from $T_0$ to the temperature $T_b$ at the boundary between cap and solvent is obtained as

$$\Delta Q = C_f \rho_f (T_b - T_0) \Delta V. \quad (14)$$

Here $C_f$ and $\rho_f$ are the specific heat capacity and the mass density of the solvent. The fluid volume, which flows along the face region of the cap in a time interval $\Delta t$, is given by

$$\Delta V = A_\parallel v_s \Delta t = \lambda_s dv_s \Delta t \quad (15)$$

with the slip velocity $v_s$ at the considered surface area and the slip length $\lambda_s$ (see Fig. 6b). $A_\parallel$ is the area cross-section of the flowing solvent in the flow direction. Thus, one has $A_\parallel = \lambda_s d$ if $d$ is the side of $A_\perp$ being perpendicular to $v_s$. By inserting equation (15) into equation (14) one obtains

$$\Delta Q = C_f \rho_f (T_b - T_0) \lambda_s dv_s \Delta t \quad (16)$$

and finally for the heat flux according to equation (13)

$$j_Q = C_f \rho_f \frac{\lambda_s}{h} v_s (T_b - T_0). \quad (17)$$

If we assume that the heat transport in the solvent occurs mainly because of the flow connected with the slip velocity, we can equate equation (17) with equation (12), which leads to the condition

$$T'_b(x) = -\frac{C_f \rho_f \lambda_s v_s(x)}{\kappa h}(T_b(x) - T_0) \quad (18)$$

at the interface between the cap and the solvent. In order to apply this condition at both sides of the particle cap in the effectively 1D model, we consider a reference slip velocity $v_0$ at the left boundary and a slip velocity $v_0 + b\sigma I'$ at the right boundary. While $\sigma$ is the particle diameter, the factor $b$ stems from the approach $v_s(x) = \bar{v} + bI(x)$. Thus, after some calculations one obtains for the constants $B_1$ and $B_2$ in equation (11)

$$B_2 = B_1 \frac{\omega - K_1 v_0}{\omega + K_1 v_0} + \frac{\alpha(I' - K_1 v_0 I_0)}{\gamma(\omega + K_1 v_0)} \quad (19)$$

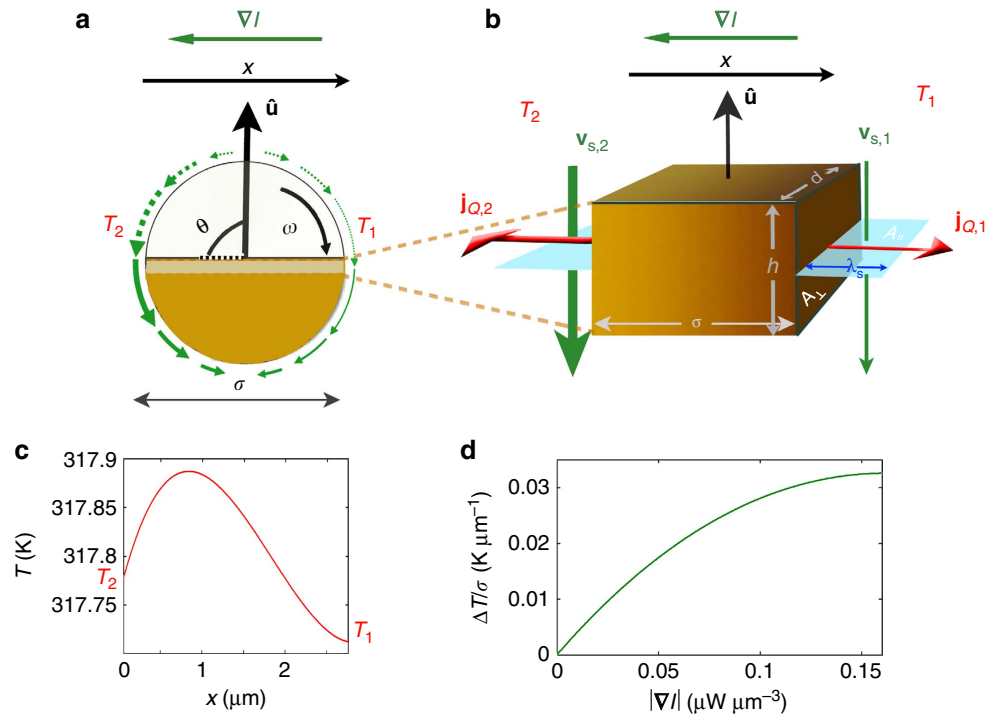

**Figure 6 | Saturation of the aligning torque.** (**a**) Schematic of an active Janus particle in a non-uniform light field with gradient $\nabla I$. The inhomogeneous illumination leads to different temperatures $T_2 > T_1$ at the two sides of the particle cap, which is approximated by the brighter region in the middle of the particle. The resulting asymmetric slip velocity profile (indicated by green arrows) induces an angular velocity $\omega$. (**b**) Close-up view of the particle cap in the effectively one-dimensional model. As the local slip velocity $v_s$ increases with higher illumination intensity, the advective coupling between the solvent and the heat flux $j_Q$ through the particle surface $A_\perp = dh$ leads to a higher flux at the left side of the particle ($j_{Q,2} > j_{Q,1}$). (**c**) Resulting temperature profile $T(x)$ inside the particle cap and (**d**) corresponding temperature gradient between the two sides of the particle as a function of the intensity gradient according to equation (27). This reduced temperature difference leads to the observed saturation behaviour of the phoretic torque. The values of the various input parameters for the theory correspond to the experimental data or are obtained from literature[61,62], respectively ($C_c = 500$ J kg$^{-1}$ K$^{-1}$, $C_f = 4,200$ J kg$^{-1}$ K$^{-1}$, $\rho_c = 2 \times 10^3$ kg m$^{-3}$, $\rho_f = 0.99 \times 10^3$ kg m$^{-3}$, $\alpha = 1.5$ K m$^2$ J$^{-1}$, $\kappa = 0.4$ W K$^{-1}$ m$^{-1}$, $\lambda_s = 180$ nm, $h = 20$ nm, $\gamma = 1.1 \times 10^5$ s$^{-1}$, $b = 10^{-9}$ m$^3$ W$^{-1}$ s$^{-1}$, $I_0 = 1$ µW µm$^{-2}$, $v_0 = 12$ µm s$^{-1}$).

with

$$K_1 = \frac{C_f \rho_f \lambda_s v_s}{\kappa h} \quad (20)$$

and

$$B_1 = \frac{\alpha(\beta_0 + \beta_1 I' + \beta_2 I'^2)}{\gamma(\delta_0 + \delta_1 I')}, \quad (21)$$

where

$$\beta_0 = \frac{(K_1^2 v_0^2 I_0 - \omega K_1 v_0 I_0)e^{-\omega\sigma}}{\omega + K_1 v_0} - K_1 v_0 I_0, \quad (22)$$

$$\beta_1 = \frac{(K_1^2 v_0 I_0 b\sigma - K_1 v_0 + \omega)e^{-\omega\sigma}}{\omega + K_1 v_0} - K_1 I_0 b\sigma - K_1 v_0 \sigma - 1, \quad (23)$$

$$\beta_2 = \frac{-K_1 b\sigma e^{-\omega\sigma}}{\omega + K_1 v_0} - K_1 b\sigma^2, \quad (24)$$

$$\delta_0 = \frac{\omega - K_1 v_0}{\omega + K_1 v_0}(K_1 v_0 - \omega)e^{-\omega\sigma} + (K_1 v_0 + \omega)e^{\omega\sigma} \quad (25)$$

and

$$\delta_1 = K_1 b\sigma \left( \frac{\omega - K_1 v_0}{\omega + K_1 v_0} e^{-\omega\sigma} + e^{\omega\sigma} \right). \quad (26)$$

The resulting temperature profile is visualized in Fig. 6c. Finally, the effective temperature gradient between the two sides of the particle is obtained as

$$\mathrm{grad}_T(I') = \frac{1}{\sigma}(T(\sigma) - T(0)) \equiv \frac{\Delta T}{\sigma}$$
$$= \frac{1}{\sigma}\left\{ B_1(I')\left[e^{\omega\sigma} - 1 + \frac{\omega - K_1 v_0}{\omega + K_1 v_0}(e^{-\omega\sigma} - 1)\right] + \left[\frac{\alpha(e^{-\omega\sigma} - 1)}{\gamma(\omega + K_1 v_0)} + \frac{\alpha}{\gamma}\sigma\right]I' \right.$$
$$\left. - \frac{\alpha(e^{-\omega\sigma} - 1)}{\gamma(\omega + K_1 v_0)} K_1 v_0 I_0 \right\}. \quad (27)$$

As shown in Fig. 6d, this gradient is reduced as compared with the applied illumination intensity, which leads to the saturation of the aligning torque observed in the experiments. Instead of the complicated result based on equations (19)–(27), for the numerical simulations we use the expression in equation (9), where the saturation is approximated by an exponential function. As the constants $C_1$ and $C_2$ are determined from experimental data, this also takes possible additional nonlinear effects into account.

**Displacement probability distribution function.** Equations (2) and (7) also allow for a calculation of the probability distribution function $P(x)$ of particle displacements. This was done in order to obtain the numerical data for Fig. 1, where the velocity profile $v_p(x) = \tilde{v}(1 + x/a)$ with $\tilde{v} = 2\,\mu m\,s^{-1}$ and $a = 10^4\,\mu m\,s^{-1}$ was used, and where $\omega(x,\varphi)$ was set to zero. The presented probability distributions are based on $10^6$ particle trajectories starting at $x=0$ with initial orientation in the positive $y$ direction. Although, after long times, the right tails of the distributions are more pronounced than the left ones (because of the larger velocities) and the mean slightly shifts to the right, no net particle flux through $x=0$ occurs (Fig. 1b). This changes for a non-zero angular velocity $\omega(x,\varphi)$, which leads to a systematic drift of the whole distribution function. (Note that in our experiments the aligning torque and a corresponding angular velocity are always present. The reference simulation visualized in Fig. 1 is only meant to illustrate that a mere position-dependent particle motility is not sufficient to create a net particle flux as observed in the experiments.)

**Aligning torque without saturation.** In order to illustrate the crucial role of the saturation of the phoretic torque, in the following, we show explicitly that for an angular velocity $\omega(x,\varphi)$ being proportional to the velocity gradient $v'_p(x)$, that is, for a linear function $f(v'_p(x)) = Av'_p(x)$ with a constant $A$ in equation (8), indeed no net particle current can occur in the limit of small noise. In that case, the equations of motion for the $x$ position and the angle $\varphi$ reduce to

$$\dot{x}(t) = v_p(x(t))\cos\varphi(t), \quad (28)$$

$$\dot{\varphi}(t) = Av_p(x(t))v'_p(x(t))\sin\varphi(t). \quad (29)$$

By using the relation $\frac{dv_p}{dx} = \frac{1}{\dot{x}}\frac{dv_p}{dt}$ and inserting equation (28), from equation (29) one obtains

$$\dot{\varphi}(t) = A\,v'_p(x(t))\tan\varphi(t). \quad (30)$$

Solving this differential equation in $\varphi$ gives

$$\sin\varphi(t) = Ce^{Av_p(x(t))}, \quad (31)$$

and thus

$$\varphi(t) = 2\pi k + \arcsin\left(Ce^{Av_p(x(t))}\right)$$
$$\text{or } \varphi(t) = 2\pi k + \pi - \arcsin\left(Ce^{Av_p(x(t))}\right) \text{ with } k \in \mathbb{Z}. \quad (32)$$

$C$ is a constant of integration, which depends on the initial angle $\varphi_0$ and the initial velocity $v_{p,0}$ and is determined by $C = \exp(-Av_{p,0})\sin\varphi_0$. The result in equation (32) provides a direct relation between the position-dependent velocity $v_p(x)$ and the particle orientation. This means that a particle with certain initial conditions will always have the same orientation once it reaches a position with a certain velocity $v_p(x)$, independent of the specific shape of the velocity profile. To illustrate that this always leads to a vanishing particle current, we now consider a particle starting in the minimum of a periodic velocity profile $v_p(x)$ (Fig. 7a). (For the example velocity profile in Fig. 7, we chose a sum of two trigonometric functions. However, the same line of argument holds for any other periodic profile as well.) If the initial angle $\varphi_0$ is zero or $\pi$, the particle will travel to the right or to the left with constant orientation and the same mean velocity as can be directly seen from equations (28) and (29). For $0 < \varphi_0 < \pi/2$ the particle is exposed to the aligning torque as soon as it enters the region with positive slope of $v_p(x)$. Now, two situations can be distinguished. If the particle reaches the orientation $\varphi = \pi/2$ before arriving at the maximum of the velocity profile (see magenta and green curves in Fig. 7), its direction of motion is reversed so that it moves back to the minimum of the profile, where it has the orientation $\pi - \varphi_0$ according to equation (32). Subsequently, the particle enters the region with negative slope of the velocity profile (see Fig. 7). Independent of the specific shape of the profile (even if there is a substructure with additional local minima and maxima as in the example in Fig. 7), the particle reaches the position of the same maximum velocity before it returns. This velocity $v_{p,max}$ is determined by the initial conditions via

$$v_{p,max} = v_{p,0} - \frac{1}{A}\ln(\sin\varphi_0). \quad (33)$$

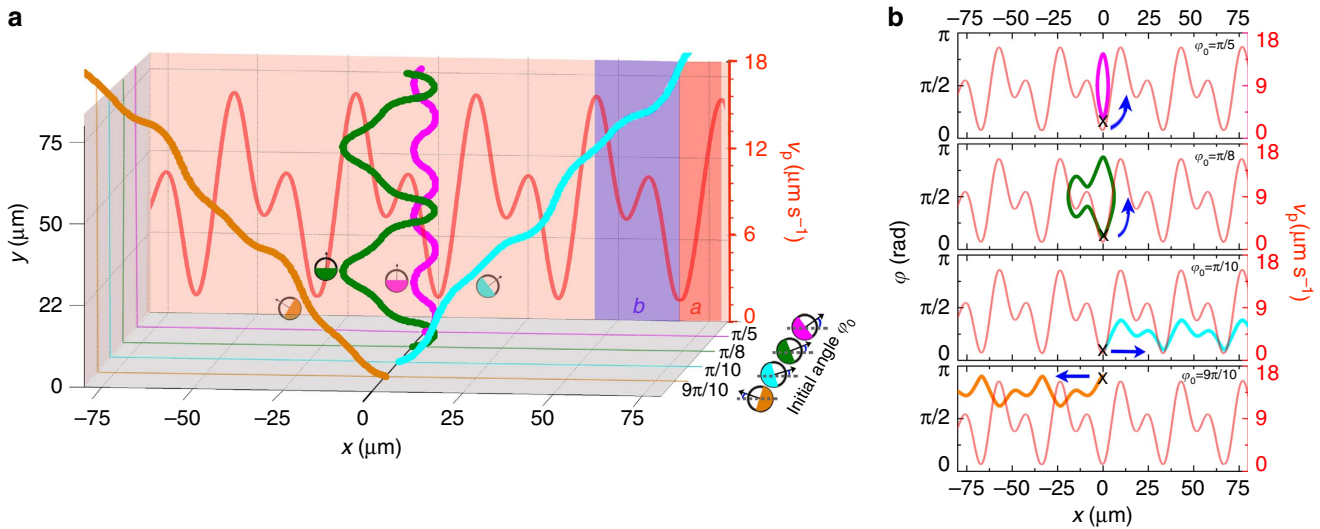

**Figure 7 | Particle dynamics without saturation of the aligning torque.** (**a**) Calculated trajectories of active Janus particles in a periodic asymmetric velocity profile $v_p(x)$ for different initial angles $\varphi_0$. Different from the experimental situation, here a linear relation between the angular velocity and the gradient of the translational velocity is considered. Depending on the initial orientation, particles either perform an oscillating motion in a valley of the velocity profile (green and magenta trajectories) or they periodically move in positive or negative $x$ direction (orange and cyan trajectories). (Note that the green trajectory is scaled by a factor of 2/5 in the $y$ direction for reasons of presentation.) In the absence of saturation, the probabilities and the mean velocities of particles moving to the left and to the right are the same. Thus, no net particle current occurs. (**b**) Particle orientation $\varphi$ versus position $x$ for the trajectories shown in **a**. Black crosses represent the initial positions, and the arrows indicate the evolution with time.

On the other hand, if the particle arrives at the maximum of the velocity profile before it is rotated to an angle $\varphi = \pi/2$ (see cyan curve in Fig. 7), it is able to pass into the next region of the profile, where the slope is negative.

Owing to this opposite sign of the gradient, the sense of rotation of the particle is reversed. Thus, when arriving at the next minimum of the profile, the particle has its initial orientation $\varphi_0$ again, and this behaviour continues periodically (Fig. 7b).

While particles trapped in one period of the velocity profile (see magenta and green trajectories in Fig. 7) can obviously not contribute to any net particle current, for particles being able to cross the peak of the profile, it has to be shown that the net current is still zero after averaging over the initial orientations of the particles. For symmetry reasons and with the result in equation (32) it is clear from equation (32) that a particle with initial orientation $\pi - \varphi_0$ moves continuously to the left, given that a particle with initial orientation $\varphi_0$ travels to the right through the periodic velocity profile. Thus, it only has to be shown that the mean velocity is the same in the two cases. For that purpose it is sufficient to calculate the time a particle with initial orientation $\varphi_0$ needs to cross region $a$ in Fig. 7 from left to right and compare it with the time a particle arriving at the minimum of the profile with orientation $\pi - \varphi_0$ needs to cross the same region from right to left. By using equation (28) and the result in equation (32), it can be shown that both crossing times are the same and are given by

$$T_{cross} = \int_{x_{start}}^{x_{end}} \frac{1}{\dot{x}} dx = \int_{x_{min}}^{x_{max}} \frac{1}{v_p(x)\cos\left(\arcsin\left(Ce^{Av_p(x)}\right)\right)} dx. \quad (34)$$

Thus, for a linear relation between the aligning torque and the gradient of the velocity profile, in a periodic profile no net particle current can occur in the limit of zero noise. In corresponding simulations with finite noise, we only obtained a marginal drift because of the Brownian motion of the active particles. However, this drift is several orders of magnitude smaller than that observed in the experiments. Therefore, the decisive aspect for the occurrence of a rectified particle motion is the existence of the saturation of the diffusiophoretic torque.

**Data availability.** The data that support the findings of this study are available from the corresponding author upon request.

### Acknowledgements

This work was supported by the German Research Foundation (DFG) through the priority programme SPP 1726 on microswimmers (H.L. and C.B.) and by the ERC Advanced Grant INTERCOCOS (Grant No. 267499). We thank R. Gomez-Solano for fruitful discussions and H.-J. Kümmerer, C. Mayer and U. Rau for their technical support.

### Author contributions

All authors designed the research, analysed the data and wrote the paper. C.L. carried out the experiments and B.t.H. performed the analytical calculations and the simulations.

### Additional information

**Supplementary Information** accompanies this paper at http://www.nature.com/naturecommunications

**Competing financial interests:** The authors declare no competing financial interests.

**Reprints and permission** information is available online at http://npg.nature.com/reprintsandpermissions/

**How to cite this article:** Lozano, C. *et al.* Phototaxis of synthetic microswimmers in optical landscapes. *Nat. Commun.* 7:12828 doi: 10.1038/ncomms12828 (2016).

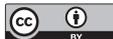